\documentclass[prl,twocolumn,superscriptaddress,showpacs,preprintnumbers,amsmath,amssymb]{revtex4-1}
\usepackage{amsmath}
\usepackage{graphicx}
\usepackage{epstopdf} 
\usepackage{dcolumn}
\usepackage{bm}
\usepackage{color}
\usepackage{enumerate}
\usepackage{mathrsfs}
\usepackage{url}
\usepackage{float}
\usepackage{makecell}
\usepackage{longtable,booktabs}
\usepackage{hyperref}

\begin{document}
\bibliographystyle{apsrev4-1} 

\title{Effect of deep-defect excitation on mechanical energy dissipation of single-crystal diamond}

\author{Huanying Sun}
\affiliation{Quantum Physics and Quantum Information Division, Beijing Computational Science Research Center, Beijing 100193, China}
\affiliation{Research Center for Materials Center, National Institute for Materials Science, Namiki 1-1, Tsukuba, Ibaraki 305-0044, Japan}

\author{Liwen Sang}
\affiliation{International Center for Materials Nanoarchitectonics (MANA), National Institute for Materials Science, Namiki 1-1, Tsukuba, Ibaraki 305-0044, Japan}

\author{Haihua Wu}
\affiliation{Research Center for Materials Center, National Institute for Materials Science, Namiki 1-1, Tsukuba, Ibaraki 305-0044, Japan}

\author{Zilong Zhang}
\affiliation{Research Center for Materials Center, National Institute for Materials Science, Namiki 1-1, Tsukuba, Ibaraki 305-0044, Japan}

\author{Tokuyuki Teraji}
\affiliation{Research Center for Materials Center, National Institute for Materials Science, Namiki 1-1, Tsukuba, Ibaraki 305-0044, Japan}

\author{Tie-Fu Li}
\affiliation{Institute of Microelectronics, Tsinghua National Laboratory of Information
Science and Technology, Tsinghua University, Beijing 100084, China}
\affiliation{Frontier Science Center for Quantum Information, Beijing 100084, China}

\author{J. Q. You}
\affiliation{Interdisciplinary Center of Quantum Information and Zhejiang Province Key Laboratory of Quantum Technology and Device, Department of Physics and State Key Laboratory of Modern Optical Instrumentation, \\ Zhejiang University, Hangzhou 310027, China}

\author{Masata Toda}
\affiliation{Graduate School of Engineering, Tohoku University, Sendai, Miyagi 980-8579, Japan}

\author{Satoshi Koizumi}
\affiliation{Research Center for Materials Center, National Institute for Materials Science, Namiki 1-1, Tsukuba, Ibaraki 305-0044, Japan}

\author{Meiyong Liao}
\email [Corresponding email:]{Meiyong.LIAO@nims.go.jp}
\affiliation{Research Center for Materials Center, National Institute for Materials Science, Namiki 1-1, Tsukuba, Ibaraki 305-0044, Japan}

\date{\today}
\begin{abstract}
The ultra-wide bandgap of diamond distinguished it from other semiconductors, in that all known defects have deep energy levels that are inactive at room temperature. Here, we present the effect of deep defects on the mechanical energy dissipation of single-crystal diamond experimentally and theoretically up to 973 K. Energy dissipation is found to increase with temperature and exhibits local maxima due to the interaction between phonons and the deep defects activated at specific temperatures. A two-level model with deep energies is proposed to well explain the energy dissipation at elevated temperatures. It is evident that the removal of boron impurities can substantially increase the quality factor of room-temperature diamond mechanical resonators. The deep-energy nature of nitrogen bestows single-crystal diamond with outstanding low-intrinsic energy dissipation in mechanical resonators at room temperature or above.
\end{abstract}
\maketitle
Recent developments in micro-electromechanical system (MEMS) and nano-electromechanical system (NEMS) resonators have created opportunities for ultrasensitive sensors for mass~\cite{Chaste_2012_NatureNano,Sage_2015_NC,Sage_2018_NC}, force~\cite{Tang_2019_MNano,Mansouri_2019_MNano}, single molecule~\cite{Hanay_2012_NatNano,Puller_2013_PRL}, and  quantum  spin detection~\cite{Rugar_2004_Nature,Barson_NanoL_2017}. Among the many materials for MEMS/NEMS, diamond has attracted intense interest owing to its excellent mechanical properties and high-temperature stability~\cite{Williams_2011_DRM,Najar_2014_APL,Khanaliloo_2015_PRX,Regan_2020_AM,Mcskimin_1972_JAP}. In particular, the coupling of mechanical resonators with nitrogen-vacancy presents a promising vision for quantum sensing~\cite{Preeti_2014_NC,Nori_2016_PRL}. Resonators with high frequency and high quality factor ($Q$) are in demand for precise measurement and quantum control. Diamond mechanical resonators have been reported with resonance frequencies above the gigahertz range~\cite{Gaidarzhy_2007_APL} and $Q$ factors as high as $10^6$~\cite{Tao_2014_NC}. Dissipative mechanisms, such as clamping losses, multiple-material dissipation, surface losses, thermoelastic dissipation (TED), mechanical defects (MD), and quantum dissipation, can determine the $Q$ factor of a mechanical resonator. With regard to intrinsic energy dissipation, in addition to TED, defects are also expected to determine the ultimate $Q$ factors of the resonators.
\par
For narrow-bandgap semiconductors, dopants can be fully ionized at room temperature (RT), reducing the $Q$ factors even below RT~\cite{Mohanty_2002_PRB,Houston_2002_APL,Gysin_2004_PRB}. Diamond has an ultra wide bandgap (UWBG) of 5.5 eV. Defects in diamond usually have deep energy natures~\cite{Sohn_2015_APL,Goss_1996_PRL,Jones_1996_APL}, which are less active at room temperature. To date, there have been no reports on the effect of deep defects on the energy dissipation of diamond. An understanding of these effects is required for enhancing ultimate device performance.
\par
In this Letter, the energy dissipation in single-crystal diamond (SCD) mechanical resonators was investigated from RT to 973 K. It is revealed that the $Q$ factors overall decrease with increasing  temperature. However, two peaks appear near 400 K and 900 K. A two-level model is proposed to explain the $Q$ factors at high temperatures. The defects corresponding to the $Q$ factors at 400 K and 900 K have deep ionization energies near 0.3 eV and 0.9 eV, respectively. From this result, we speculate that single-crystal diamond is an ideal candidate for intrinsically high $Q$ factor mechanical resonators, because the only unintentional nitrogen defect is inactive at RT. This observation is contrary to semiconductor devices, in which shallow dopants are needed for enough carriers but induce mechanical dissipation, and deep defects have the negative effect of reducing electron mobility~\cite{Sze_2008_Book, Romero_2020_PRA}.
\begin{figure}
\centering
\includegraphics[width=2.75in]{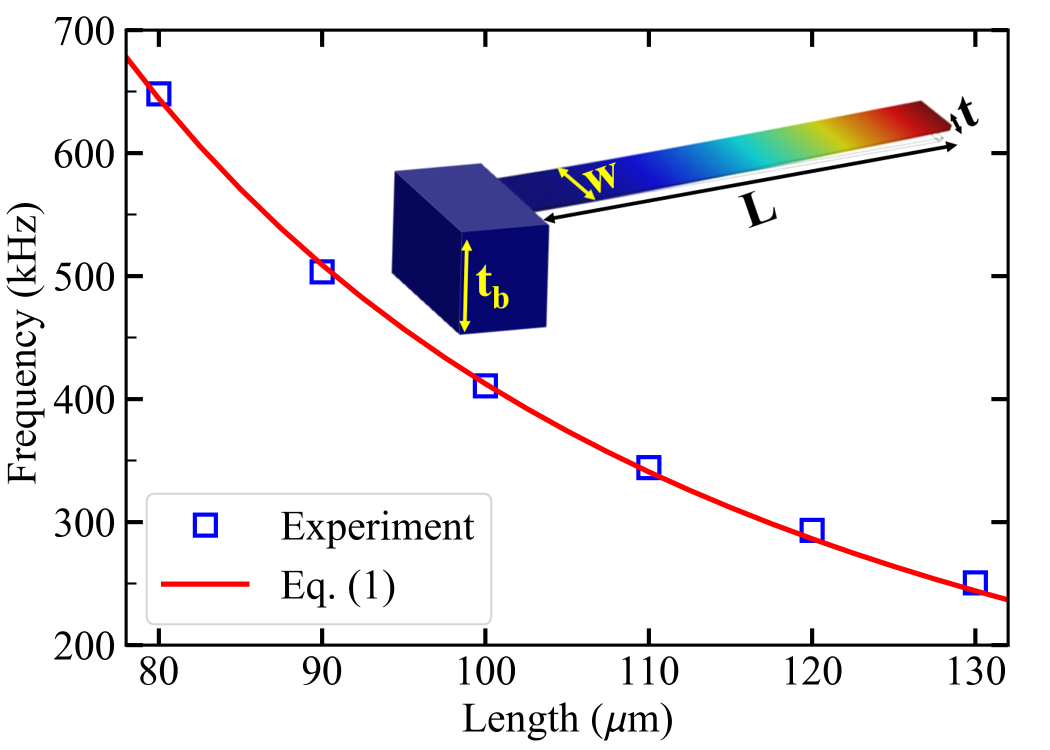}
\caption{\label{fig:1}(Color online.) Dependence of resonance frequency on cantilever length of sample DS1 at room temperature. The blue squares are the experimental measurements, and the red line  represents the calculated values based on Eq.~(\ref{eq:1}) with $t=1.44 ~\mu$m, $E = 1100$ GPa, and $\rho = 3.5~\rm{g/cm^3}$. The inset represents the schematic of a single SCD cantilever.}
\end{figure}
\par
The SCD cantilevers were fabricated by a smart-cut method as described in~\cite{MeiyongLiao_2010_AM,ZilongZhang_2019_Carbon}, which has the structure of SCD-on-SCD. The SCD cantilevers of diamond sample 1 (DS1)  were processed in oxygen ambient at 773 K to remove the ion-irradiated defects at the bottom of the cantilevers~\cite{HaihuaWu_2018_PRM}. The dimensions of the cantilevers are as follows: thickness $t = 1.44 ~\mu$m and length $L$ from $80 ~\mu$m to $140~\mu$m, and  width $w$ of 6 and 12~$\mu$m.  The inset in Fig.~\ref{fig:1} shows a schematic of an SCD cantilever. To obtain the resonance frequency spectra, the SCD cantilevers were actuated by a radio-frequency signal applied to a probe placed aside the SCD cantilevers. The out-of-plane resonance frequencies of the cantilevers were detected with a laser Doppler vibrometer~\cite{ZilongZhang_2019_Carbon}. The samples were measured in a high vacuum with a pressure of approximately $10^{-4}$ Pa to eliminate air damping. The measurements were conducted from RT to 1000 K in steps of 25 K, which was controlled by a Lake Shore Model 335 temperature controller. To exclude the effect of the surface adsorbates on the resonance frequencies and $Q$ factors, the SCD cantilevers were annealed at 1000 K for 1 h before the detailed measurements (Fig.~S1)~\cite{Supp_Infor}. The $Q$ factor is calculated as $Q = \omega/\Delta\omega$. The measured resonance frequency spectra were fitted on the basis of the Lorentzian line shape (Fig.~S2)~\cite{Thomas_2019_OE,Born_2013_Book, Supp_Infor}. Here, $\omega$ is the resonance frequency and $\Delta\omega$ is the full width at half maximum (FWHM) of the resonance peak. To preclude the effect of thermal drift on the $Q$ factors, we measured the frequency spectra back and forth, until there was no change in the resonance peak.
\par
First, the length-dependent resonance frequencies of the SCD cantilevers were utilized to rule out the stress effect on the resonance frequency and $Q$ factors~\cite{Scott_2006_JAP}. We confirmed that the out-of-plane resonance frequency of the cantilevers can be described in the following form~\cite{Weaver_1990_Book}, as shown in Fig.~\ref{fig:1}:
\begin{equation}\label{eq:1}
\begin{aligned}
\omega_n=K_n^2\frac{t}{L^2}\sqrt{\frac{E}{12\rho}}
\end{aligned}
\end{equation}
where $K_n$ is the vibration mode index, with $K_1 = 1.8751, K_2 = 4.6941, K_3 = 7.8548, ..., K_n = (2n-1)\pi/2$, and $L$ and $t$ are the length and thickness of the cantilever, respectively. For SCD, the Young's modulus $E \approx 1100$ GPa and the density of mass $\rho = 3.5~ \rm{g/cm^3}$ at RT. In addition, we fit the length-dependent resonance frequency by considering the stress in the cantilevers by~\cite{Lonkar_2013_APL,Ikehara_2000_JMM}
\begin{equation}\label{eq:2}
\begin{aligned}
\omega_n(\sigma)=\omega_n\sqrt{1-\frac{\sigma A L^2}{K_n^2EI}}
\end{aligned}
\end{equation}
where $\sigma$ is the magnitude of the uniaxial compressive stress along the length of the beam, $I=wt^3/12$ is the bending moment, and  $A$ is the cross-sectional area. It is shown in Table~S1 that the stress in the SCD cantilevers is negligible~\cite{Supp_Infor}.
\par
\begin{figure}
\centering
\includegraphics[width=2.75in]{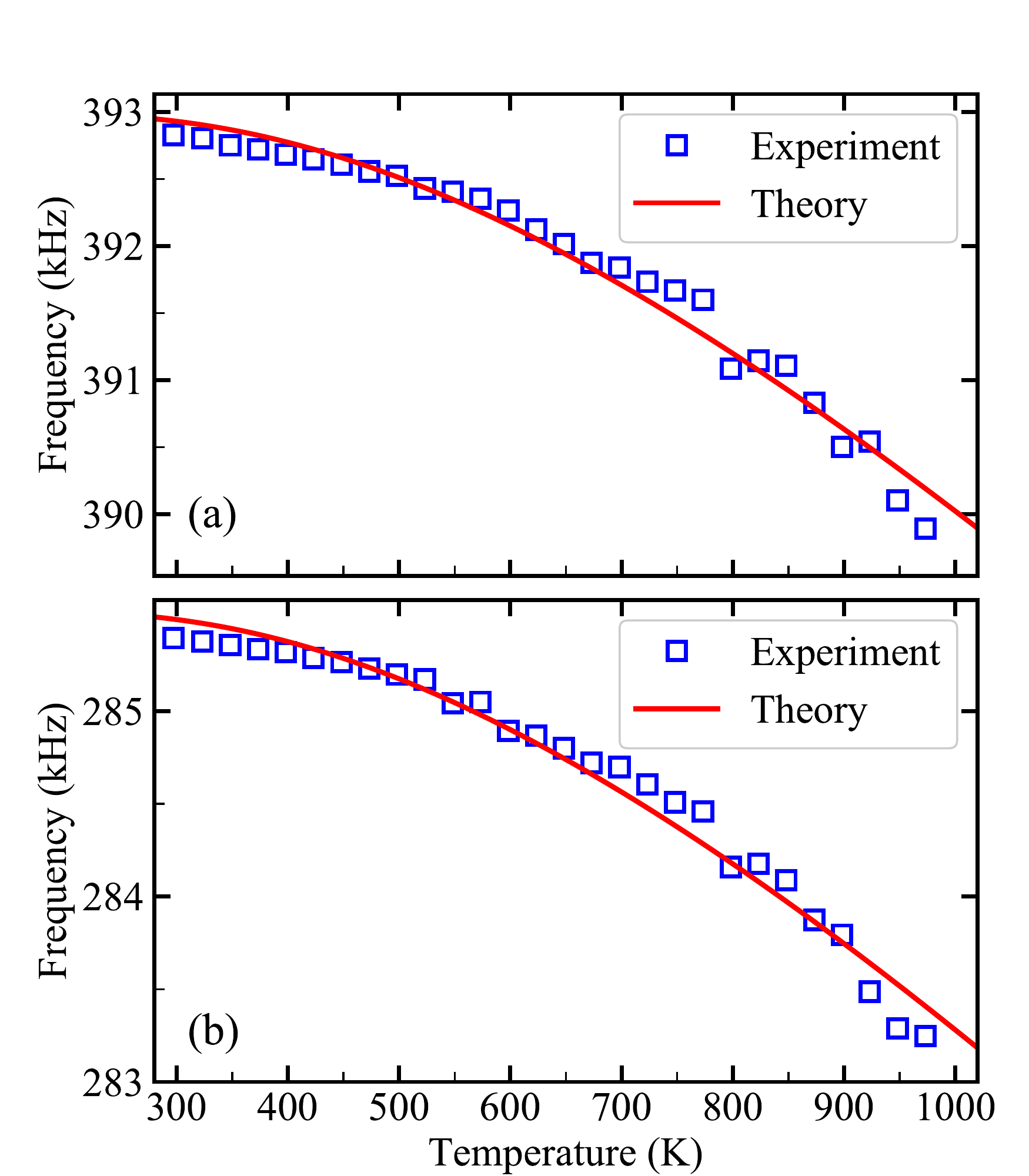}
\caption{\label{fig:2} (Color online.) Dependence of resonance frequency on temperature of the SCD cantilevers in sample DS1 with (a) $L = 100~\mu$m of cantilever III-4 and (b) $L = 120~\mu$m of cantilever IV-3. The blue squares represent the experimental data and red lines represent the theoretical fitting results modeled by Eq.~(\ref{eq:3}). The fitting parameters and the dimensions of each cantilever are listed in Table~\ref{table:1}.}
\end{figure}

Second, we investigated the temperature-dependent resonance frequency to extract the Young's modulus as a function of temperature for subsequent energy dissipation analysis. The resonance frequency of the SCD cantilevers (III-4 and IV-3 of sample DS1) as a function of temperature is shown in Fig.~\ref{fig:2}. The exponential decrease in the experimental resonance frequency with temperature is modeled by the temperature-dependent Young's modulus~\cite{Wachtman_1961_PR}:
\begin{equation}\label{eq:3}
\begin{aligned}
E(T)=E_0-CT{\rm exp}{\left(-\frac{T_0}{T}\right)}
\end{aligned}
\end{equation}
where $E_0$ is the  Young's modulus at 0 K. $C$ is a constant independent of temperature, and $T_0$  is a characteristic temperature related to the SCD Debye temperature with $T_0~=~\Theta_D/2$. Anderson theoretically obtained  a similar formula based on the change in specific heat with temperature~\cite{Anderson_1966_PR}. The fitting result of the resonance frequency with temperature is shown by the solid lines in Fig.~\ref{fig:2}, where the fitting was conducted by combining Eqs.~(\ref{eq:1})-(\ref{eq:3}), the parameters for the fittings are listed in Table~\ref{table:1}. The variation in resonance frequency with temperature further confirms  the negligible stress in the cantilevers.  A similar tendency of resonance frequency $vs$ temperature was also observed for other cantilevers, as shown in Fig.~S3(a) and Fig.~S4~\cite{Supp_Infor}.
\begin{table}
\caption{\label{table:1} The optimal fitting parameters for temperature-dependent resonant frequency for two SCD cantilevers in sample DS1. Here, Young's modulus $E_0$ and the Debye temperature $\Theta_D$ are set as 1100 GPa and 2230 K, respectively~\cite{Vogelgesang_1996_PRB}.}
\begin{tabular}
{p{1.65cm}<{\centering}p{1.54cm}<{\centering}p{1.54cm}<{\centering}p{1.54cm}<{\centering}p{1.54cm}<{\centering}p{1.54cm} }
\hline \hline
& Length & Width & $C$ &$\sigma$   \rule{0pt}{0.45cm}\\
Sample 1 &($\mu$m) &($\mu$m)  &(MPa)&(MPa)\\
\hline
III-4 &100 & 12 &45.16  &6.14 \\
IV-3 &120 & 12 &51.79  &0.27 \\
\hline \hline
\end{tabular}
\end{table}
\par
\begin{figure}
\centering
\includegraphics[width=2.75in]{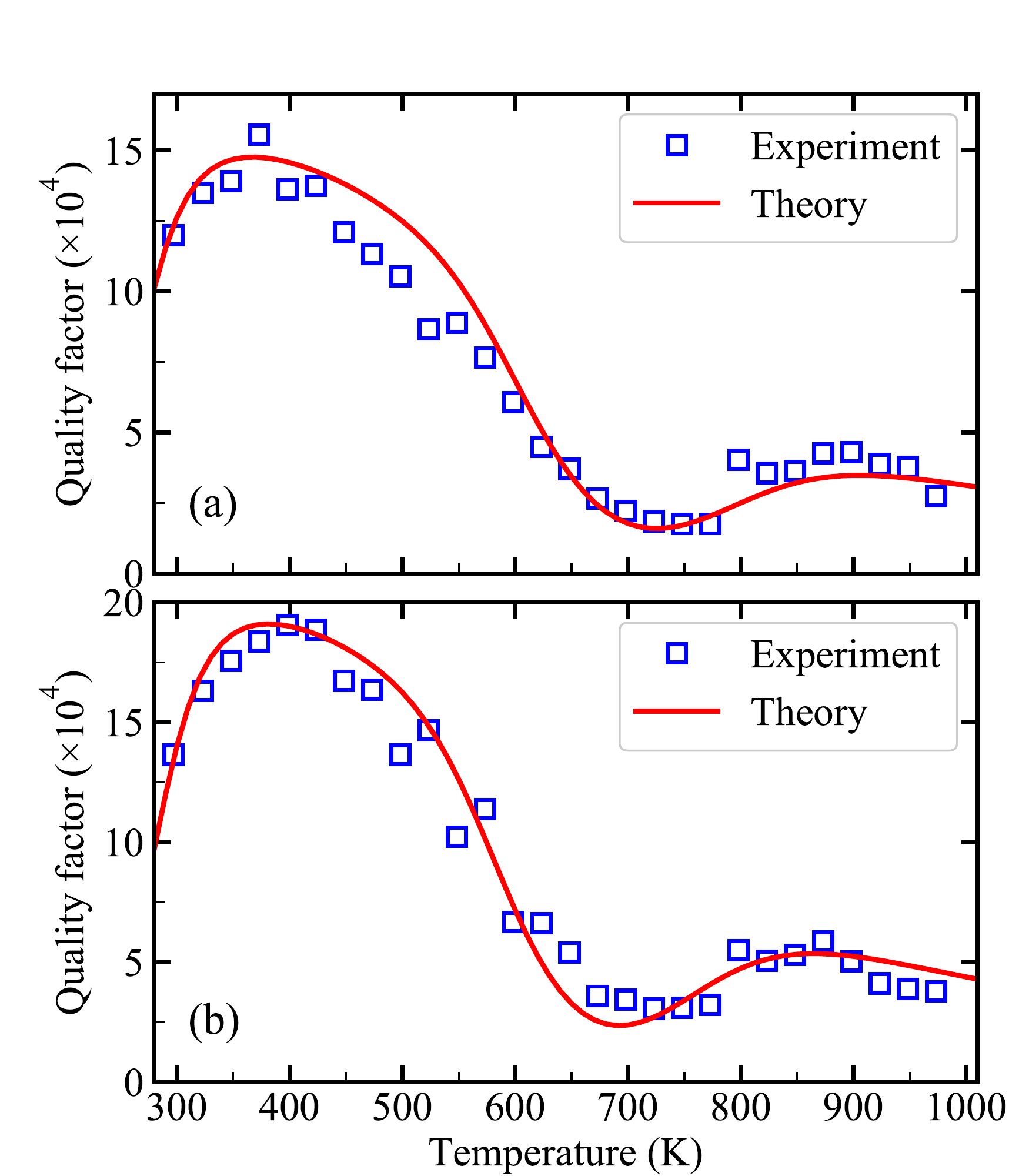}
\caption{\label{fig:3} (Color online.) Dependence of quality factor on temperature for (a) cantilever III-4 and (b) cantilever IV-3. The blue squares represent the experimental data and red lines represent the theoretical calculation. The parameters used in the theoretical calculations are listed in Table S3~\cite{Supp_Infor}.}
\end{figure}

The quality factor is defined as the ratio of the stored energy in the resonator to the dissipated energy per cycle of vibration~\cite{Pozar_2009_Book,Schmid_2016_Book}. When there is no strain in the mechanical resonator, a higher $Q$ factor means a lower rate of damping or energy loss~\cite{Schmid_2016_Book}. In the presence of a large strain, the strain may induce dissipation dilution and markedly improve the $Q$ factor~\cite{Scott_2006_JAP,Scott_2007_NanoL}, where the increase in $Q$ factor is due to the increase in energy stored rather than the decrease in dissipation. In this case, the mechanical dissipation rate $\gamma = \omega/Q$ is used~\cite{Lukin_2010_PRL,Verhagen_2012_Nature}. In the present SCD cantilevers, the strain is negligible, as indicated by the length-dependent resonance frequency in Fig.~\ref{fig:1}. In Fig.~\ref{fig:3}, the temperature-dependent $Q$ factor is shown from RT to 973 K for the same cantilevers in Fig.~\ref{fig:2}. It is revealed that the $Q$ factor on average decreases with increasing temperature. The $Q$ factors are greater than 10,000 over the full range of temperature and the maximum value is close to 200,000 near 400 K. Two  maxima near 400 K and 900 K  are observed. Similar behavior of $Q$ factors $\emph{vs}$ temperature was also observed for other cantilevers in the same sample DS1 (Fig.~S3(b) and Fig.~S5). On the other hand, the resonance frequency also decreases with temperature, however, with the variation is less than 1 percent even up to 1000 K because of the decrease in the Young's modulus of diamond. To further exclude the effects of strain and temperature-dependent resonance frequency,  the curves of $Q/\omega$ {\emph{vs}} temperature are plotted in Fig.~S6. It can be seen that the same two maxima appear near 400 K and 900 K. For simplification, here, the $Q$ factors were approximately utilized to describe the energy dissipation~\cite{Maximilian_2018_APL}.
\par
To interpret the temperature-dependent $Q$ factors, we analyzed the dominant energy dissipation mechanisms by excluding others. We assumed that the damping loss from air or surrounding gas $Q_{air}^{-1}$~\cite{Schmid_2016_Book} was negligible because the samples were in high vacuum.  The bulk loss from grain boundaries, dislocations, and other crystal phases were also assumed to be negligible owing to the SCD nature~\cite{MeiyongLiao_2010_AM}. The remaining losses were then considered to be dominated by clamping
$Q_{\emph clamp}^{-1}$,  surface $Q_{\emph surface}^{-1}$, TED $Q_{\emph TED}^{-1}$, and MD dissipation $Q_{\emph MD}^{-1}$, where the TED and MD losses are temperature-dependent. From the temperature-dependent $Q$ factors, we assumed two types of mechanical defects in our experiments. Therefore, we adopted the following model~\cite{Mohanty_2013_PR,MeiyongLiao_2014_APL}:
\begin{equation}\label{eq:4}
Q_{\emph tot}^{-1}=Q_{\emph clamp}^{-1}+Q_{\emph surface}^{-1}+Q_{\emph TED}^{-1}+Q_{\emph MD1}^{-1}+Q_{\emph MD2}^{-1}
\end{equation}
As shown in Eq.~(S3) and Eq.~(S4) in the Supplementary Information~\cite{Supp_Infor} and discussion herein, the clamping loss and surface effect show almost no dependence on temperature, and thus cannot be the mechanisms for the observed maxima near 400 K and 900 K shown in Fig.~\ref{fig:3}.
\par
Here, we primarily discuss the energy loss mechanisms of TED and MD, with a focus on the defects-related MD mechanism. Crystal defects, such as substitutional impurities and interstitial motion, reconfigure between equilibrium and metastable states if the dynamic strain fields are changed by external conditions~\cite{Mohanty_2013_PR,Hutchinson_2004_APL,Czaplewski_2005_JAP,Yildiz_2019_NatureM}.  The dissipation originating from  mechanical defects is expressed as
\begin{eqnarray}
\label{eq:7} Q_{\emph MD}^{-1}&=&\delta \frac{\omega\tau_{\emph MD}}{1+(\omega\tau_{\emph MD})^2}\\
\label{eq:8} \tau_{\emph MD}&=&\tau^{-1}_{\emph MD0}{\rm exp}\left(-\frac{E_A}{k_BT}\right)
\end{eqnarray}
where $\delta$ is a unitless constant related to defect concentration, $\tau_{\emph MD}$ is the defect relaxation time, and $\omega$ is the cantilever resonant frequency. $\tau_{\emph MD0}$ is the characteristic atomic vibration period on the order of $10^{-13}$ s. $E_A$ is the activation energy of the defect. The mechanical defect dissipation reaches its maximum when $\omega \tau_{\emph MD}=1$ and results in a Debye peak.
\par
\begin{figure}
\centering
\includegraphics[width=2.75in]{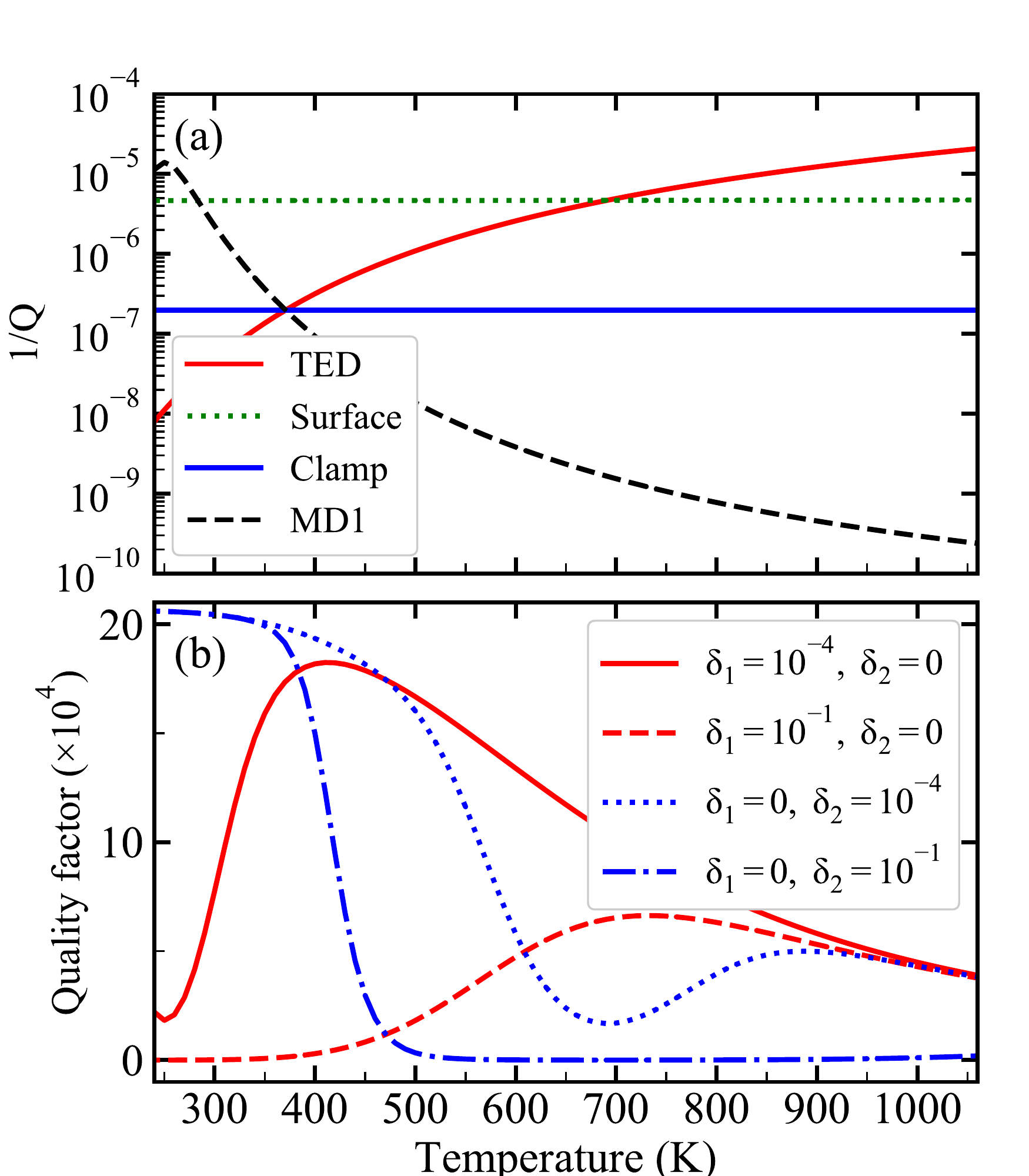}
\caption{\label{fig:4} (Color online.) (a) Variation of each dissipation term in Eq. (4) with temperature. MD1 is the mechanical dissipation of defect D1 and the dissipation of defect D2 is zero and not shown. Here, $\delta_1 = 10^{-4}$ and $\delta_2 = 0$. (b) Calculated $Q$ factors $vs$ temperature with different defect concentrations. Each line represents only one kind of defect being activated. The red lines are for D1 with excitation energy 0.33 eV. The blue lines are for D2 with excitation of 0.91 eV. The other calculation parameters are the same as those of cantilever IV-3 in Table S3.}
\end{figure}

The theoretical total dissipation  is obtained by calculating each term in Eq.~(\ref{eq:4}) and compared with the experimental data, as shown in Fig.~\ref{fig:3}. The corresponding calculation parameters shown in Table~S3 refer to the literature~\cite{Inyushkin_2018_PRB,Olsen_1993_PRB,LanHua_Wei_1993_PRL}. To clearly determine the temperature-dependent dissipation mechanism, the theoretical curve of each mechanism in Eq.~(\ref{eq:4}) is plotted in Fig.~\ref{fig:4}(a). The numerical results show that the clamping and surface losses change little with temperature rising. The increased TED loss with increasing temperature in Fig.~\ref{fig:4}(a) is solely due to the reduction in the thermal conductivity, which is responsible for the $Q$ factor reduction on average (Fig.~S7). Therefore, only the mechanical defect dissipation contributes to the observed maxima at approximately 400 K and 900 K in the relationship between $Q$ factor and temperature (Fig.~\ref{fig:3}). The simulation by the mechanical defects model reveals that the thermal activation energy at approximately 400 K is 0.31 -- 0.33 eV, named D1 later. This activation energy value is similar to that of boron acceptors in diamond~\cite{Prins_1988_PRB,Sandhu_1989_APL}. The defect contributing to the maximum of the $Q$ factor near 900 K was simulated to have a thermal activation energy of 0.91 -- 0.92 eV, named D2 later. The D2 defect is likely related to the nitrogen impurities~\cite{Hagiwara_1988_JPSJ,Tzeng2017_Book}. In addition, two other SCD samples without boron (DS2) and with high boron concentration (DS3) were measured for comparison. For both samples, the damaged substrate layer was not removed, accounting for the relatively low $Q$ factors. For the cantilevers of sample DS2 without boron dopants, the maximum $Q$ factor near 900 K was observed, as shown in Fig.~S9~\cite{Supp_Infor}, but the peak at approximately 400 K was not obvious. In another sample, DS3, which was deliberately doped with boron at a concentration of $5\times10^{18}/\rm{cm^3}$, the peak near 400 K was obvious, as shown in Fig.~S10~\cite{Supp_Infor}. In sample DS3, the reason the declining $Q$-factor trend was not observed at high temperatures may be the compensation effect of boron defects.
\par
From the simulation, we expected that the  defect density would strongly affect the $Q$ factors at different temperatures. By calculating each energy dissipation term in Eq.~(\ref{eq:4}), as shown in Fig.~\ref{fig:4}(a), the total $Q$  factors were obtained, as shown in Fig.~\ref{fig:4}(b). To understand the dissipation of the mechanical defects, we considered the extreme cases with one type of defect dominating. We changed the defect density-related parameters $\delta_1$ and $\delta_2$ and used the other parameters listed in Table S3. In Fig.~\ref{fig:4}(b), the red lines represent the case where only defect D1 is activated. The solid and dashed lines represents the defect concentrations of $\delta_1 = 10^{-4}$ and $\delta_1 =  10^{-1}$, respectively. The blue lines represent the case where only defect D2 is activated. The dash-dotted line and dotted line represent the defects concentrations of $\delta_2 = 10^{-4}$ and $\delta_2 = 10^{-1}$, respectively. It is revealed that if the concentration of D1 is sufficiently high, then the $Q$ factor is low even at RT. In contrast, D2 reduces the $Q$ factor only at high temperatures, even at a rather high concentration. In Fig.~\ref{fig:4}(a), we calculated the mechanical defect dissipation with the special case of $\delta_1 = 10^{-4}$ (with D1) and $\delta_2 = 0$ (without D2) by using the same parameters as those of cantilever IV-3 in Table~S3. It is revealed that the mechanical defect dissipation of D1 decreases with increasing temperature and only the first maximum originating from boron appears near RT. The dissipation of D1 directly leads to the maximum $Q$ factor near 400 K, as shown in Fig.~\ref{fig:4}(b). Similarly, the maximum energy dissipation of D2 appears near 700 K if we take $\delta_1 = 0$ (without D1) and $\delta_2 = 10^{-4}$ (with D2). Because boron (D1) usually  can be avoided and nitrogen (D2) is the only unintentional impurity in diamond, diamond has obvious advantages for high $Q$-factor resonators because of the deep-defect nature of nitrogen. Because of the low intrinsic mechanical losses in diamond, by using strain-diluted dissipation, SCD resonators with ultra-high $Q$ factors of more than $10^8$ might be achieved \cite{Scott_2006_JAP,Tsaturyan_2017_NaNoTech,Ghadimi_2018_Science,Fedorov_2020_PRL}.
\par
In conclusion, we investigated the mechanical energy dissipation of SCD from RT to 1000 K. It is shown that the energy dissipation increases with increasing temperature. Two maxima in the $Q$ factor were observed near 400 K and 900 K. It is theoretically argued that these maxima are consistent with the excitation of deep defects at elevated temperatures, and the defect concentration strongly affects the overall mechanical energy dissipation. Considering the features of deep defects in diamond, diamond MEMS resonators have the advantage of high $Q$ factors at elevated temperatures. This experiment also provides a promising method for controlling the $Q$ factor by defect engineering in other UWBG semiconductors.
\par
This work was supported by Science Challenge Project (No. TZ2018003), JSPS KAKENHI (Grant No.20H02212, 15H03999, 26220903), JST-PRESTO (Grant No. JPMJPR19I7), and Nanotechnology Platform projects sponsored by the Ministry of Education, Culture, Sports, and Technology (MEXT) in Japan, National Key Research Development Program of China (No. 2016YFA0301200), NSAF (No. U1930402), and BAQIS Research Program (No. Y18G27). H. Sun also gratefully thanked financial support from China Scholarship Council (No. 201904890013).
%
\end{document}